\documentclass[twocolumn,trackchanges] {aastex631}
\shorttitle{Blue-shifted Fe XXV He$\alpha$ line on IM Peg}
\shortauthors{Inoue et al.}
\graphicspath{{./}{figures/}}
\usepackage{threeparttable}
\usepackage{color}
\usepackage{enumerate}
\usepackage{multirow}

\begin{document}

\title{High-velocity blue-shifted Fe XXV He$\alpha$ line during a superflare of the RS CVn-type star IM Peg}

\author[0000-0003-3085-304X]{Shun Inoue}
\affiliation{Department of Physics, Kyoto University, Kitashirakawa-Oiwake-cho, Sakyo-ku, Kyoto, 606-8502, Japan}
\email{inoue@cr.scphys.kyoto-u.ac.jp}

\author[0000-0002-0207-9010]{Wataru Buz Iwakiri}
\affiliation{International Center for Hadron Astrophysics, Chiba University, Inage-ku, Chiba, 263-8522 Japan}

\author[0000-0003-1244-3100]{Teruaki Enoto}
\affiliation{Department of Physics, Kyoto University, Kitashirakawa-Oiwake-cho, Sakyo-ku, Kyoto, 606-8502, Japan}
\affiliation{RIKEN Cluster for Pioneering Research, 2-1 Hirosawa, Wako, Saitama, 351-0198, Japan}

\author[0000-0003-1518-2188]{Hiroyuki Uchida}
\affiliation{Department of Physics, Kyoto University, Kitashirakawa-Oiwake-cho, Sakyo-ku, Kyoto, 606-8502, Japan}

\author[0000-0002-3133-9053]{Miki Kurihara}
\affiliation{Department of Astronomy, Graduate School of Science, The University of Tokyo, 7-3-1 Hongo, Bunkyo-ku, Tokyo 113-0033, Japan}
\affiliation{Japan Aerospace Exploration Agency, Institute of Space and Astronautical Science, Chuo-ku, Sagamihara, Kanagawa 252-5210, Japan}

\author[0000-0002-9184-5556]{Masahiro Tsujimoto}
\affiliation{Japan Aerospace Exploration Agency, Institute of Space and Astronautical Science, Chuo-ku, Sagamihara, Kanagawa 252-5210, Japan}

\author[0000-0002-0412-0849]{Yuta Notsu}
\affiliation{Laboratory for Atmospheric and Space Physics, University of Colorado Boulder, 3665 Discovery Drive, Boulder, CO 80303, USA}
\affiliation{National Solar Observatory, 3665 Discovery Drive, Boulder, CO 80303, USA}


\author[0000-0001-7515-2779]{Kenji Hamaguchi}
\affiliation{Astrophysics Science Division, NASA’s Goddard Space Flight Center, Greenbelt, MD 20771, USA}
\affiliation{Department of Physics, University of Maryland, Baltimore County, Baltimore, MD, USA}

\author[0000-0001-7115-2819]{Keith Gendreau}
\affiliation{Astrophysics Science Division, NASA’s Goddard Space Flight Center, Greenbelt, MD 20771, USA}

\author{Zaven Arzoumanian}
\affiliation{Astrophysics Science Division, NASA’s Goddard Space Flight Center, Greenbelt, MD 20771, USA}

\author[0000-0002-5504-4903]{Takeshi Go Tsuru}
\affiliation{Department of Physics, Kyoto University, Kitashirakawa-Oiwake-cho, Sakyo-ku, Kyoto, 606-8502, Japan}



\begin{abstract}
Monitor of All-sky X-ray Image (MAXI) detected a superflare, releasing $5\times 10^{37}$ erg in 2$-$10 keV, of the RS CVn-type star IM Peg at 10:41 UT on 2023 July 23 with its Gas Slit Camera (GSC; 2$-$30 keV).
We conducted X-ray follow-up observations of the superflare with \textit{Neutron Star Interior Composition ExploreR} (\textit{NICER}; 0.2$-$12 keV) starting at 16:52 UT on July 23 until 06:00 UT on August 2. 
\textit{NICER} X-ray spectra clearly showed emission lines of the Fe XXV He$\alpha$ and Fe XXVI Ly$\alpha$ for $\sim 1.5$ days since the MAXI detection. The Fe XXV He$\alpha$ line was blue-shifted with its maximum Doppler velocity reaching $-2200 \pm 600$ $\mathrm{km \: s^{-1}}$, suggesting an upward-moving plasma during the flare, such as a coronal mass ejection (CME) and/or chromospheric evaporation. This is the first case that the Fe XXV He$\alpha$ line is blue-shifted during a stellar flare and its velocity overwhelmingly exceeds the escape velocity of the star ($-230$ $\mathrm{km \: s^{-1}}$).
One hour before the most pronounced blueshift detection, a signature of reheating the flare plasma was observed.
We discuss the origin of the blueshift, a CME or high-velocity chromospheric evaporation.
\end{abstract}



\section{Introduction} \label{sec:intro}
Stellar magnetic activity, such as flares and coronal mass ejections (CMEs), and their effects on exoplanets have recently attracted wide attention \citep[e.g.,][]{Airapetian_2020}.
One particular observational advance over the past decade is the increase of measurements of prominence eruptions, the initial interval of stellar CMEs.
Optical spectroscopic observations revealed “blueshifts” or “blue asymmetries” of chromospheric lines, which are attributed to prominence eruptions \citep[][]{Vida_2016, Flores_2017, Fuhrmeister_2018, Honda_2018, Vida_2019, Muheki_2020a, Muheki_2020b, Koller_2021, Maehara_2021, Lu_2022, Namekata_2021, Inoue_2023, Notsu_2024, Namekata_2023, Inoue_2024}. 
However, the information of the chromospheric line is limited in the initial interval of mass ejection \citep{Sinha_2019}.
Optical observations alone cannot determine whether the erupted prominence eventually develops into a CME, escaping from the star.
Thus, it is essential to observe X-ray lines, which have the information of the motion of CMEs.

\epsscale{1.1}
\begin{figure*}[t]
 \begin{center}
  \plotone{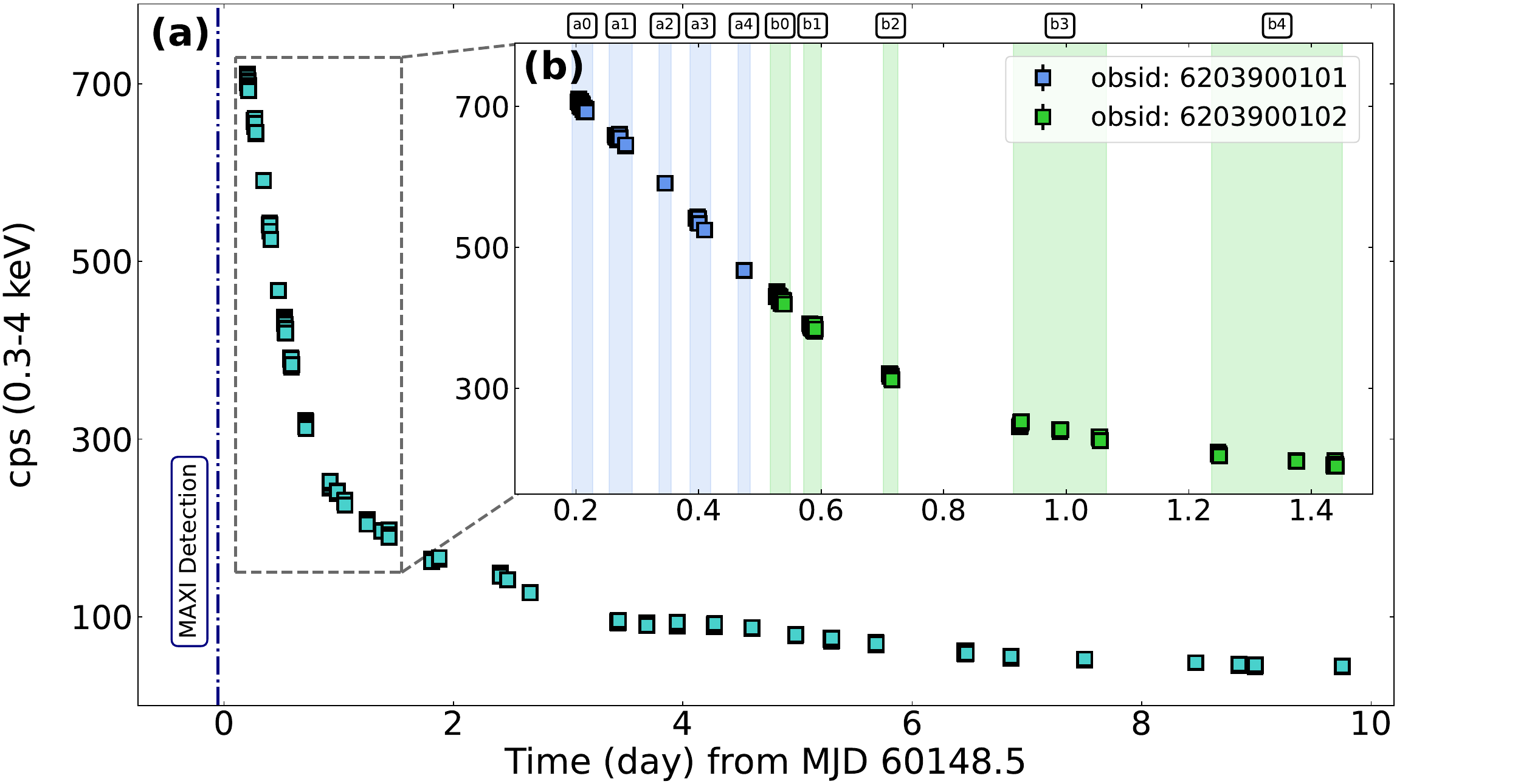}
  \caption{(a) 64 s binned $0.3-4$ keV count rate (cps; $\mathrm{counts \: s^{-1}}$) during all \textit{NICER} observations of IM Peg. The time zero of MJD 60148.5 corresponds to 12:00 UT 2023 July 23. The one standard deviation statistical error bars are smaller than the symbol size. The navy vertical dashdot line indicates the time when MAXI detected the flare ($\sim - 0.05$ day). (b) Enlarged light curve during the first two ObsIDs 6203900101 (blue) and 6203900102 (green). Interval numbers in our spectral analysis are shown above.
  \label{fig:light_curve}}
 \end{center}
\end{figure*}

\epsscale{1.1}
\begin{figure*}[]
 \begin{center}
  \plotone{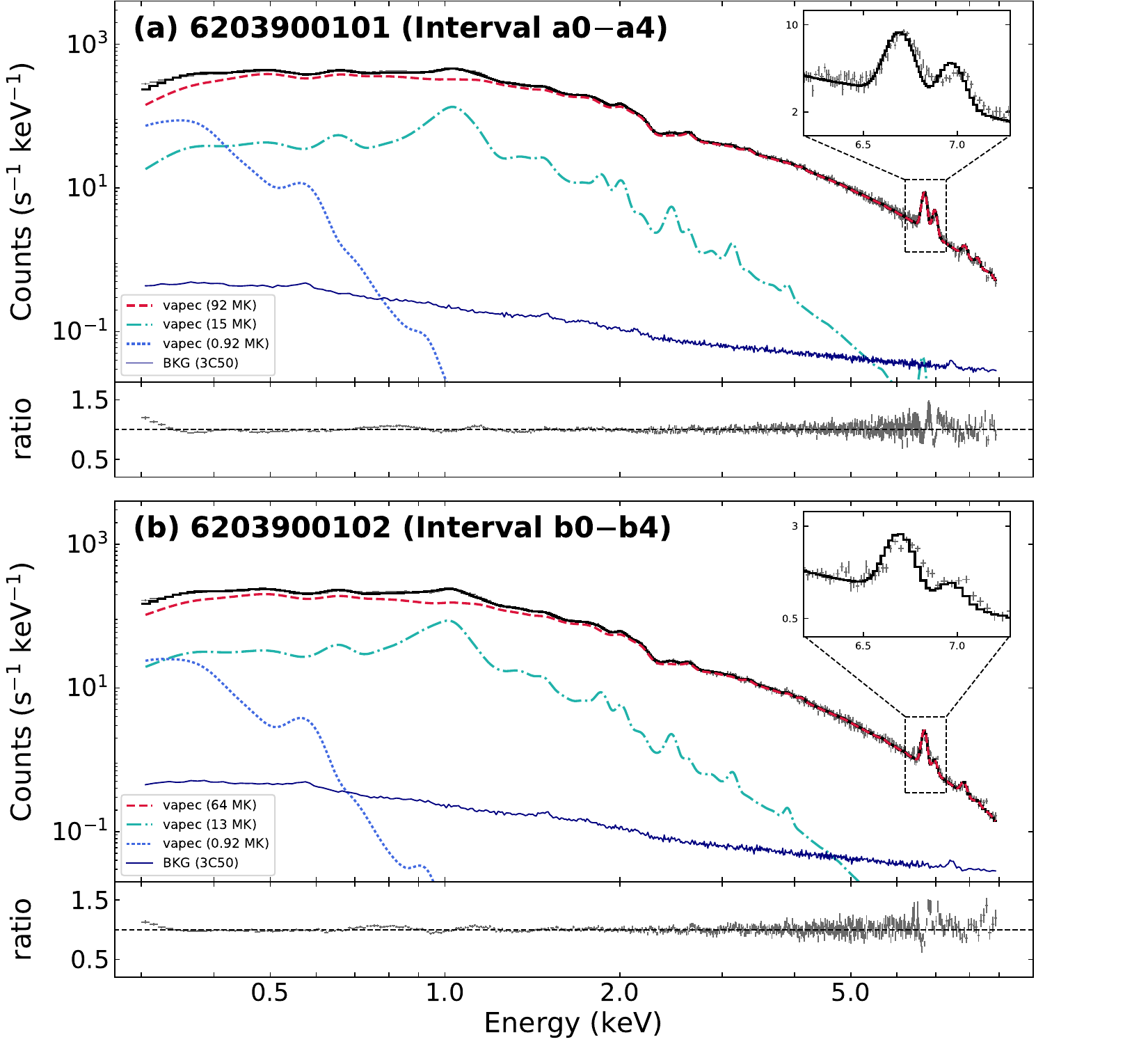}
  \caption{Background-subtracted $0.3-9$ keV \textit{NICER} spectra of IM Peg duing the interval a0$-$a4 (panel a, ObsID 6203900101) and b0$-$b4 (panel b, ObsID 6203900102). The best-fit model using the three-temperature CIE components (\texttt{vapec}) are shown by the solid black line, composed of 92 MK (red dashed), 15 MK (green dashdot), and 0.92 MK (blue dotted) for panel a and 64 MK (red dashed), 13 MK (green dashdot), and 0.92 MK (blue dotted) for panel b.
  The navy solid lines show the background spectra calculated with the 3C50 model.
  The right upper inset panels show the enlarged spectra around the iron emission lines.
  The lower panels are the ratio of the observed data relative to the best-fit model.
  \label{fig:wideband}}
 \end{center}
\end{figure*}

So far, there are only two X-ray studies of detecting the blueshifts caused by the plasma flow during a stellar flare \citep{Argiroffi_2019, Chen_2022}.
Both observations were conducted by High-Energy Transmission Grating \citep[HETG:][]{Canizares_2005} on Chandra X-ray Observatory \citep{Weisskopf_2000}.
The stars of which \citet{Argiroffi_2019} and \citet{Chen_2022} reported blueshifts were the G type star HR9024 and M type star EV Lac, respectively.
The reported velocity of blueshifts is $\sim - 100$ $\mathrm{km \: s^{-1}}$, below the escape velocity at the star surface.
Both \citet{Argiroffi_2019} and \citet{Chen_2022} discussed the formation temperatures of the lines and discussed whether the blueshifts are attributed to a CME or chromospheric evaporation \citep{Fisher_1985}.
However, since these HETG studies investigated only the lines formed at temperatures below $20$ MK ($\sim 1.7$ keV), it was not known whether high-temperature ($>50$ MK) lines such as the Fe XXV He$\alpha$ and Fe XXVI Ly$\alpha$ are blueshifted during flares. 
On the solar flares such high-temperature ($>50$ MK) plasma is thought to be formed above the top of a flare loop, and is related to the particle acceleration \citep{Caspi_2010}.



Large flares with its energy exceeding $10^{35} \: \mathrm{erg}$ from magnetically active RS CVn-type stars have been observed in both X-ray and visible light \citep[e.g.,][]{Tsuru_1989, Tsuboi_2016, Sasaki_2021, Inoue_2023}. \citet{Inoue_2023} reported a prominence eruption during a flare releasing $7\times10^{35}$ erg in the RS CVn type star V1355 Orionis, of which outward velocity was measured to be $1600$ $\mathrm{km \: s^{-1}}$ by $\mathrm{H\alpha}$ observations. 
On the other hand, no blueshifted X-ray lines have been reported despite many X-ray observations of RS CVn-type stars \citep[e.g.,][]{Testa_2004}.
The good photon statistics of large flares from RS CVn-type stars could allow us to statistically investigate a blueshift of the Fe XXV He$\alpha$ line.

In this letter, we report the X-ray observation of a superflare from the RS CVn-type star IM Peg with \textit{Neutron Star Interior Composition ExploreR} \citep[\textit{NICER};][]{2016SPIE.9905E..1HG}. 
The large effective area of \textit{NICER} \citep[600 $\mathrm{cm}^{2}$ at 6 keV;][]{Gendreau_2012, Arzoumanian_2014} allows us to conduct time-resolved spectral analysis during flares.
We describe observation and data reduction (Section \ref{sec:observation_reduction}), results and discussion (Section \ref{sec:analysis_results}), and conclusion (Section \ref{sec:conclusions}).
In this paper, the error ranges indicate 90 \% confidence level unless otherwise indicated.

\epsscale{0.995}
\begin{figure*}[]
  \plotone{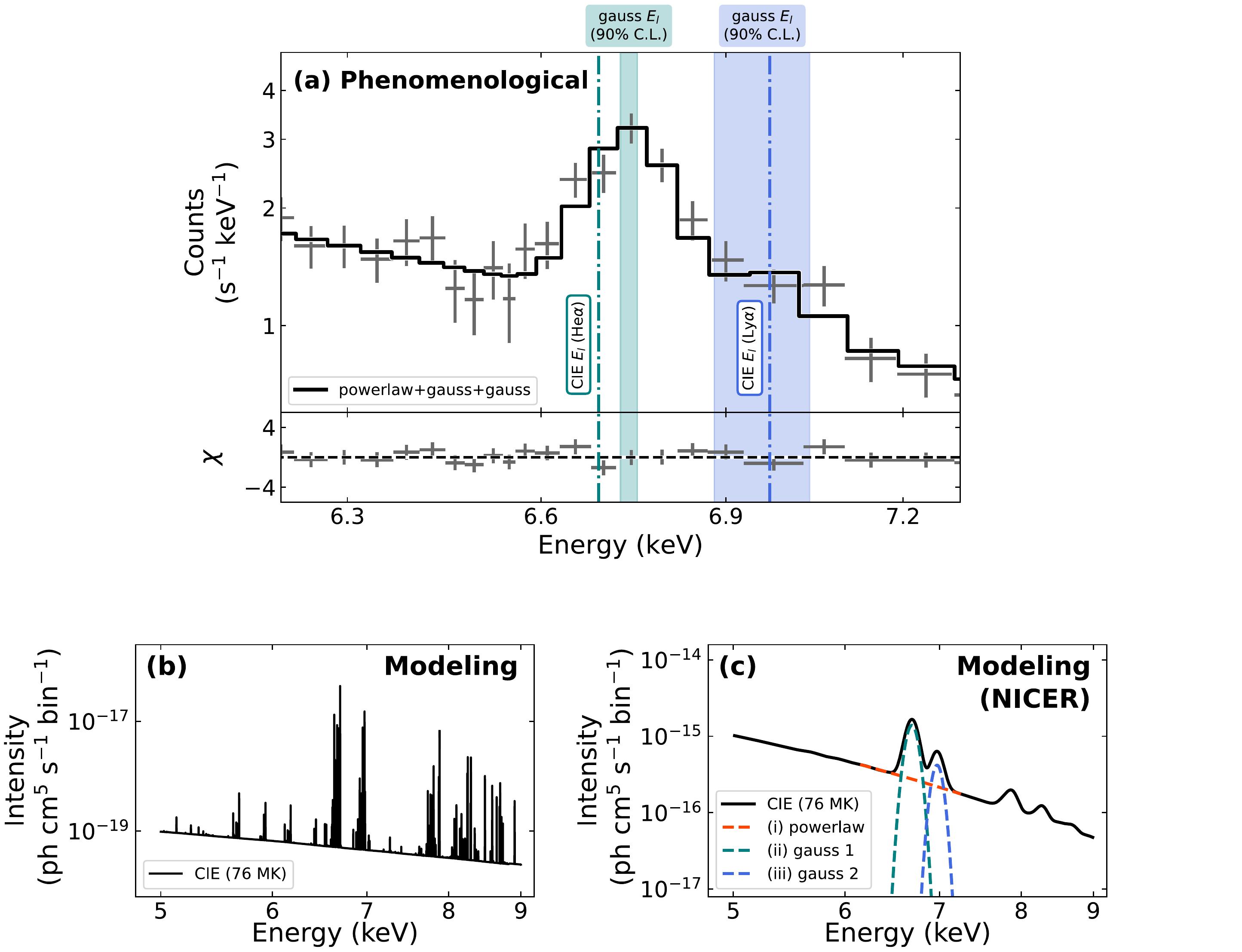}
  \caption{
  (a) The interval b1 spectrum fitted with a phenomenological model (\texttt{powerlaw+gauss+gauss}). The vertical green and blue dashdot lines indicate the line center of He$\alpha$ and Ly$\alpha$ emission lines at the rest frame in the CIE model, respectively. The green and blue shaded areas show the $90\%$ error range of the line center of the two gaussians. (b) Theoretically expected CIE spectrum during the interval b1 at the electron temperature of $T_{e}^{\mathrm{b1}} \sim 76$ MK calculated with \texttt{PyAtomDB}. (c) Same spectrum as panel (b) folded by the \textit{NICER} response function.
  \label{fig:spectra}}
\end{figure*}

\begin{table}[]
\caption{Best-fit spectral parameters of the phenomenological model during the interval b1}
\begin{center}
\begin{tabular}{cccccc}
\hline \hline
\multicolumn{6}{c}{Phenomenonological model (\texttt{powerlaw+gauss1+gauss2})}                                                                                                  \\ \hline \hline
\multirow{2}{*}{\texttt{powerlaw}}                      & $\Gamma$        & \multicolumn{4}{c}{$2.21\pm0.18$}     \\
                                                        & $K^{\mathrm{powerlaw}}$   ($\mathrm{keV^{-1} \: cm^{-2} \: s^{-1}}$)         & \multicolumn{4}{c}{$0.28\pm0.07$}     \\ \hline
\multirow{3}{*}{\texttt{gauss1}}                         & $E_{l}^{\mathrm{gauss1}}$ (keV)   & \multicolumn{4}{c}{$6.74\pm0.01$}     \\ 
                                                        & $\sigma$ (keV)  & \multicolumn{4}{c}{$0.00$ (fix)}     \\
                                                        & $K^{\mathrm{gauss1}}$ ($10^{-3} \: \mathrm{cm^{-2} \: s^{-1}}$)& \multicolumn{4}{c}{$1.25\pm0.19$}     \\ \hline              
\multirow{3}{*}{\texttt{gauss2}}                         & $E_{l}^{\mathrm{gauss2}}$ (keV)   & \multicolumn{4}{c}{$6.96\pm0.08$}     \\    
                                                        & $\sigma$ (keV)  & \multicolumn{4}{c}{$0.00$ (fix)}     \\  
                                                        & $K^{\mathrm{gauss2}}$ ($10^{-3} \: \mathrm{cm^{-2} \: s^{-1}}$)& \multicolumn{4}{c}{$0.29\pm0.14$}     \\ \hline 
\multicolumn{2}{c}{$\chi^{2}$ (d.o.f.)} & \multicolumn{4}{c}{49 (72)} \\ 
\multicolumn{2}{c}{Null hyp. prob.} & \multicolumn{4}{c}{0.98} \\ \hline \hline
\end{tabular}
\end{center}
\tablecomments{
The unit of $K^{\mathrm{powerlaw}}$ is $\mathrm{photons \: keV^{-1} \: cm^{-2} \: s^{-1}}$ at 1 keV, whereas $K^{\mathrm{gauss}}$ is the $\mathrm{total \: photons \: cm^{-2} \: s^{-1}}$ in the line. The negative velocity indicates that the spectrum is blue-shifted.}
\label{tab:fitted_parameter}
\end{table}

\section{Observation and Data Reduction} \label{sec:observation_reduction}
IM Peg is a well-known active RS CVn-type star \citep[e.g.,][]{Buzasi_1987}, of which basic parameters are summarised in Table 1 of \citet{Zellem_2010}, e.g., the stellar mass and radius of this K-type star are $M=1.8M_{\odot}$, and $R=13.3R_{\odot}$, respectively, where $M_{\odot}$ and $R_{\odot}$ are the mass and radius of the Sun, respectively. 
The corresponding escape velocity at the stellar surface is $v_{e} = \sqrt{2GM/R} = 230 \: \mathrm{km \: s^{-1}}$.
This long-period RS CVn binary consists of K2 III + G4V stars with orbital period of 24.6 d. The K2 III star has synchronized rotation and is heavily spotted and the likely source of the superflare.

Monitor of All-sky X-ray Image \citep[MAXI;][]{Matsuoka_2009} detected a flare from IM Peg at 10:41 UT on 2023 July 23.
This flare was classified as a superflare by the radiated energy of $5\times10^{37}$ erg in 2$-$10 keV \citep{Iwakiri_2023}.
We carried out follow up observations with \textit{NICER} from 16:52 UT on July 23 to 06:00 UT on August 2.

We downloaded the \textit{NICER} data of ObsIDs: 6203900101$-$6203900111 from the HEASARC archive.
We employed the standard analysis procedure.
First, we used \texttt{nicerl2} in HEASoft ver. 6.32.1 to filter and calibrate raw data using the calibration database (\texttt{CALDB}) version \texttt{xti20221001}
with two options, which 
were (a) overshoot count rate range of 0$-$5; (b) cut off rigidity greater than 1.5 $\mathrm{GeV \: c^{-1}}.$
Then, we extracted light curves from the filtered events with \texttt{xselect} and generated ObsID-averaged source and background spectra with \texttt{nicerl3-spect} and the 3C50 model \citep{Remillard_2022}. We also extracted time-resolved spectra of ObsIDs 6203900101 and 6203900102 with \texttt{nimaketime}, \texttt{niextract-event} and \texttt{nicerl3-spect}. 
We used \texttt{Xspec ver. 12.12.1} \citep{1996ASPC..101...17A} and \texttt{PyXspec ver. 2.1.0} \citep{Gordon_2021} for our spectral analyses.


\epsscale{0.84}
\begin{figure*}[]
\begin{center}
  \plotone{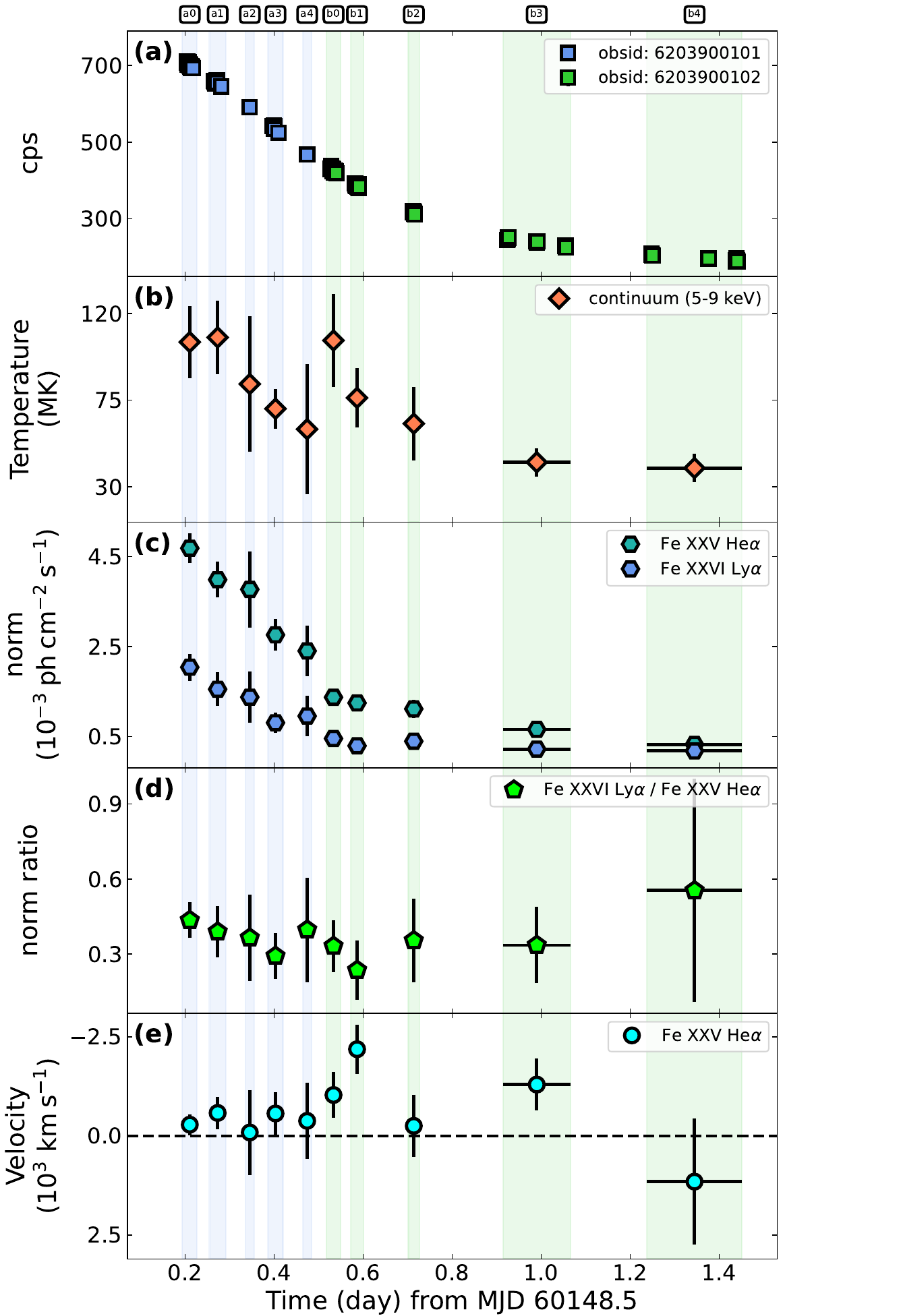}
\end{center}
  \caption{ (a) 0.3$-$4 keV light curve same as Figure \ref{fig:light_curve}b. (b) Time evolution of the electron temperature calculated from the 5$-$9 keV continuum. (c) Best-fit normalization ($K^{\mathrm{gauss1}}, \: K^{\mathrm{gauss2}}$) of the two gaussians. (d) Ly$\alpha$/He$\alpha$ ratio calculated by the two gaussian best-fit normalizations ($K^{\mathrm{gauss2}} / K^{\mathrm{gauss1}}$). (e) Doppler velocity of the line center of the Fe XXV He$\alpha$ line. The negative sign corresponds with blue-shift.
  \label{fig:parameter}}
\end{figure*}

\section{Analysis and Result}\label{sec:analysis_results}
Figure \ref{fig:light_curve} shows the $0.3-4$ keV light curve of all \textit{NICER} observations of IM Peg covering the decay phase of the flare for $\sim 10$ days.
We divided the first two ObsIDs 6203900101 and 6203900102 into good time intervals (GTIs) with their typical exposures of 500$-$1500 s. These GTIs are numbered as the interval a0$-$b4 (Figure \ref{fig:light_curve}b).
We combined the three GTIs of the interval b3 and b4 to obtain better statistics for the following spectral analysis.

First, we extracted two spectra from the interval a0$-$a4 and b0$-$b4 (Figure \ref{fig:wideband}).
We fit the wide-band (0.3$-$9 keV) spectra with a collisionally-ionized equilibrium (CIE) model composed of three temperature components (\texttt{vapec+vapec+vapec}) convolved with interstellar absorption (\texttt{tbabs}).
We tied the abundance among the three components.
The best-fit temperatures of the three components of the interval a0$-$a4 and b0$-$b4 were 92 MK, 15 MK, and 0.92 MK and 64 MK, 13 MK, and 0.92 MK, respectively.

The Fe XXV He$\alpha$ and Fe XXVI Ly$\alpha$ lines were clearly detected during the interval a0$-$a4 and b0$-$b4 (Figure \ref{fig:wideband}), as expected from the temperatures (92 MK; a0$-$a4 and 64 MK; b0$-$b4) of the hottest plasma components \citep{Smith_2001}.
The \textit{NICER} energy resolution of 137 eV at 6 keV cannot resolve the fine structures of the Fe XXV He$\alpha$ and Fe XXVI Ly$\alpha$ lines; the former consists of the resonance, intercombination, and forbidden lines at 6.60$-$6.72 keV, whereas the latter of Ly$\alpha_{1}$ and Ly$\alpha_{2}$ lines at 6.95$-$6.97 keV (Appendix \ref{appendix: Energy_table}).
Therefore, both lines are approximated by a single Gaussian function with their lab-frame line centers at $\sim 6.69$ keV (He$\alpha$) and $\sim 6.97$ keV (Ly$\alpha$).

The simple CIE model fittings show residuals around the iron line emissions (6$-$7 keV) in the ObsID-averaged spectra (interval a0$-$a4 and b0$-$b4 in Figure \ref{fig:wideband}).
The highest-temperature component of the CIE model successfully explain the high-energy continuum above 5 keV, but fails to reproduce the line profiles (the ratio panels in Figure \ref{fig:wideband}).
The enlarged spectra around the iron emission lines in Figure \ref{fig:wideband} suggest that they are blue-shifted.
To investigate the time evolution of these two iron emission lines, we further extracted eight spectra of the individual interval a0$-$b4 (see Appendix \ref{appendix: wideband}).
Then, we found the most pronounced residuals around the iron emission lines appeared in the interval b1 spectrum.
Neither the CIE nor off-CIE models can explain the 5--9 keV spectrum during the interval b1 without assuming that the iron emission lines were blue-shifted, which suggests that the plasma is moving toward us (Appendix \ref{appendix: plasma_models}).

\begin{sloppypar}
In the following analysis, we focus on the spectral fittings around the iron features in 6$-$7 keV.
In Figure \ref{fig:spectra}a, we analyzed the interval b1 spectrum with a phenomenological model (\texttt{powerlaw+gauss1+gauss2}) in narrow-band (5$-$9 keV).
The best-fit parameters of the line center of the \texttt{gauss1} and \texttt{gauss2} were $E_{l}^{\mathrm{gauss1}} = 6.74 \pm 0.01$ keV and $E_{l}^{\mathrm{gauss2}} = 6.96 \pm 0.08$ keV, respectively (Table \ref{tab:fitted_parameter}).
To investigate how far these results deviate from the lab-frame ones, we modeled the CIE spectrum (Figure \ref{fig:spectra}b) and folded it by \textit{NICER} response functions (Figure \ref{fig:spectra}c) at the electron temperature of the interval b1 with \texttt{PyAtomDB} \citep{Foster_2020}.
In this process, we used \texttt{RMF} and \texttt{ARF} files made for the interval b1 event file with \texttt{nicerl3-spect} (Section \ref{sec:observation_reduction}).
During the interval b1, the continuum fitting with \texttt{apec}, ignoring the iron line bands (i.e., 5$-$9 keV except for 6.3$-$7.4 keV), gives its electron temperature at $T_{e}^{\mathrm{b1}} = 76$ MK. 
We also fit the model spectrum (Figure \ref{fig:spectra}c) with the phenomenological model (\texttt{powerlaw+gauss1+gauss2}) and get the best-fit parameters of $E_{l}^{\mathrm{He \alpha}} (T_{e}^{\mathrm{b1}}) = 6.69$ keV and $E_{l}^{\mathrm{Ly \alpha}} (T_{e}^{\mathrm{b1}}) = 6.97$ keV. 
Given these results, the energy shifts of the Fe XXV He$\mathrm{\alpha}$ and Fe XXVI Ly$\mathrm{\alpha}$ lines are calculated as $\Delta E^{\mathrm{He \alpha}} =  E_{l}^{\mathrm{gauss1}} - E_{l}^{\mathrm{He \alpha}} (T_{e}^{\mathrm{b1}})  = 0.05 \pm 0.01 \: \mathrm{keV}$ and $\Delta E^{\mathrm{Ly \alpha}} =  E_{l}^{\mathrm{gauss1}} - E_{l}^{\mathrm{He \alpha}}(T_{e}^{\mathrm{b1}}) = - 0.01 \pm 0.08 \: \mathrm{keV}$, respectively.
The Doppler velocity of $v^{\mathrm{He \alpha}} = - c \Delta E^{\mathrm{He \alpha}} / E_{l}^{\mathrm{He \alpha}} (T_{e}^{\mathrm{b1}}) = -2200 \pm 600 \: \mathrm{km \: s^{-1}}$ (90 \% confidence level) overwhelmingly exceeded the escape velocity of IM Peg ($-230 \: \mathrm{km \: s^{-1}}$).
On the other hand, the error range of $v^{\mathrm{Ly \alpha}} = - c \Delta E^{\mathrm{Ly \alpha}} / E_{l}^{\mathrm{Ly \alpha}} (T_{e}^{\mathrm{b1}})  = 500 \pm 3500 \: \mathrm{km \: s^{-1}}$ were so large that we could not quantitively confirm a blueshift of the Fe XXVI Ly$\alpha$ line.
\end{sloppypar}

Using the same procedure, we also estimated the electron temperature and the Doppler velocity of the Fe XXV He$\alpha$ and Fe XXVI Ly$\alpha$ lines for all intervals.
Figure \ref{fig:parameter}a$-$e show the time evolution of physical parameters obtained by our narrow-band (5$-$9 keV) spectral analysis.
The electron temperature and intensities (normalizations) of two lines decrease with time, while the line intensity ratio (Fe XXVI Ly$\alpha$ / Fe XXV He$\alpha$) stay constant within errors.
These results are consistent with the superflare of the RS CVn-type star UX Ari reported in \citet{Kurihara_2024}.
One interesting point is that there was a sign of reheating only during the interval b0 (Figure \ref{fig:parameter}b).
Figure \ref{fig:parameter}e shows that there is a tendency for the line center of the Fe XXV He$\alpha$ line to be blue-shifted except for the interval b4.
The velocity has a peak value of $-2200 \pm 600 \: \mathrm{km \: s^{-1}}$ at the interval b1.
The intervals other than b1 when the blueshift was clearly identified were the intervals b0 and b3, with velocities of $-1000\pm500 \: \mathrm{km \: s^{-1}}$ and $-1300\pm650 \: \mathrm{km \: s^{-1}}$, respectively.


In order to detect the blueshift at the 50 eV level, it is also essential to check the energy calibration of the detector. 
As a reference of a relative energy shift among the \textit{NICER} observational data, we selected supernovae remnant Cas A, which has been frequently observed by \textit{NICER}. 
This source has a similar X-ray intensity as IM Peg during the flare, the prominent iron Fe XXV He$\alpha$ emission line, and is not time-varying. We extracted 605 spectra with the same exposure as each GTI of IM Peg, of which spectra are individually fitted. The 605 Fe XXV He$\alpha$ emission line center follows the Gaussian distribution with a central value of $6.602_{-0.00015}^{+0.00047}$ keV, which is a reasonable value for the Fe XXV He$\alpha$ line and within the range reported by previous satellite observations ($6.56-6.62$ keV; Appendix \ref{appendix: nicer_accuracy}). The width of the Gaussian distribution of this 605 sample is 8.7 eV (26 eV at 3$\sigma$, equivalent to $1200 \: \mathrm{km \: s^{-1}}$), which is sufficiently smaller than the velocity discussed in this paper.
We also confirmed that the light leak had little impact on the \textit{NICER} data since most of our observations (intervals a0$-$a4, b0$-$b4) were performed during an orbital night with low background (undershoot values below $\sim 10 \: \mathrm{counts \: s^{-1}}$).
Furthermore, the \textit{NICER} team has been checking time-dependent instrumental gain shifts of \textit{NICER} detectors in 2018--2021. They suggest the shift in 2023 is at most 10--15 eV near the Fe-K line energy band (private communication with Dr. Craig Markwardt as the lead of \textit{NICER} calibration activities).
This instrumental gain shift is smaller than the detected $\sim 50$ eV blueshift.

\section{Discussion and Conclusion}\label{sec:conclusions}
We presented the superflare of the RS CVn-type star IM Peg with \textit{NICER} from 16:52 UT on 2023 July 23.
During the early interval of the observations, \textit{NICER} spectra clearly showed Fe XXV He$\alpha$ and Fe XXVI Ly$\alpha$ emission lines.
The Fe XXV He$\alpha$ line was blue-shifted during the flare, which suggests that the plasma is moving toward us.
By phenomenological fitting and CIE modeling, the maximum Doppler velocity of the Fe XXV He$\alpha$ line was estimated to be $-2200\pm600 \: \mathrm{km \: s^{-1}}$.

This measurement allows us to estimate the total kinetic energy of the blue-shifted plasma, which could lead to understanding global energetics of superflares \citep[cf.][]{Emslie_2012}.
The flare loop size is estimated to be $L_{\mathrm{SY}} \sim 10^{-1} R$ using the magnetic reconnection model equations in \citet{Shibata_2002} with assuming that the temperature and emission measure of the interval a0 (Figure \ref{fig:wideband_three}a) to be the peak values.
We assumed that the coronal density is $n_{e} = 10^{10-13} \: \mathrm{cm^{-3}}$ \citep{Aschwanden_1997, Gudel_2004, Reale_2007, Sasaki_2021} in this calculation.
Then, we calculated the kinetic energy of the blueshift as $E_{\mathrm{kin}}^{\mathrm{He\alpha}} \sim L_{\mathrm{SY}}^{3} n_{0} m_{\mathrm{p}} (v^{\mathrm{He \alpha}})^{2} \sim 10^{36} \: \mathrm{erg}$, where $m_{\mathrm{p}}$ is the mass of the proton.
The estimated kinetic energy is less than a few parcent of the X-ray flare energy of $5 \times 10^{37} \: \mathrm{erg}$ \citep{Iwakiri_2023}.

There are two possible cases for interpreting the blue-shifted Fe XXV He$\alpha$ line: a CME or chromospheric evaporation.
In the CME interpretation, the $\sim - 2200$ $\mathrm{km \: s^{-1}}$ blueshifts of the Fe XXV He$\alpha$ line originate from a CME. The typical velocity of chromospheric evaporation is $\sim - 100$ $\mathrm{km \: s^{-1}}$ \citep[e.g.,][]{Doschek_1980}.
The velocity of solar CMEs \citep[$-20 \sim -3000$ $\mathrm{km \: s^{-1}}$:][]{Webb_2012} can explain the observed velocity.
The CME, whose velocity overwhelmingly exceeded the escape velocity ($\sim - 230 \: \mathrm{km \: s^{-1}}$), would certainly have flown outward from the star.
Furthermore, we see that there is a trend of acceleration from the interval a4 to b1 (Figure \ref{fig:parameter}g).
This is consistent with \citet{Argiroffi_2019}, in which they discussed a blueshift during the post-flare phase as the CME.

It is also possible that the blue-shifted emission from the CME was present during the interval a0$-$a4 as well as b1, but was not clearly observed due to the predominance of strong emission from the flare component.
In this interpretation, we can also fit the Fe XXV He$\alpha$ line with two gaussian components, the flare component and the CME component.
In the case, the velocity of the CME were estimated to be $-6100 \pm 900$ $\mathrm{km \: s^{-1}}$ for the interval a0$-$a4 and $-4400 \pm 1000$ $\mathrm{km \: s^{-1}}$ for the interval b0$-$b4, respectively (Appendix \ref{appendix: two_gaussian}).

The formation temperature ($T_{\mathrm{peak}}$) of the Fe XXV He$\alpha$ line ($T_{\mathrm{peak}}\sim$ 60 MK) is higher than that of other stellar CME candidates.
\citet{Argiroffi_2019} discussed a later blueshift of $\mathrm{O_{VIII}}$ line ($T_{\mathrm{peak}}\sim 3$ MK) on the G type star HR9024 as a CME candidate.
\citet{Chen_2022} reported upflows of 3$-$16 MK plasma accompanied by the decrease of the coronal density on the M type star EV Lac.
The temperature of solar CMEs is also on the order of a few MK \citep{Sheoran_2023}.
The discovery of the Fe XXV He$\mathrm{\alpha}$ line blueshift, which could not be investigated by the Chandra HETG observations \citep{Argiroffi_2019, Chen_2022}, means that CMEs on magnetically active stars like RS CVn-type stars consist of hot plasma ($\sim$ 60 MK).

In the chromospheric evaporation interpretation,
the reheating during the interval b0 (Figure \ref{fig:parameter}d) one hour before the most pronounced blueshift can be well understood.
From the interval a4 to b0, the electron temperature rose from 60 MK to 100 MK over the course of an hour.
For some reason, the plasma in the chromosphere may have been rapidly reheated at roughly half a day after the flare peak, causing the high-velocity chromosphere evaporation.

To conclude, this work reported the blueshifts of the Fe XXV He$\alpha$ line during a stellar flare for the first time.
Given the impact of CMEs on exoplanets \citep[e.g.,][]{Airapetian_2016}, it is important to determine whether such blueshifts are attributed to a CME or chromospheric evaporation.
To know the answer, we need to distinguish whether the line itself is blue-shifted or non-shifted and blue-shifted components are added together.
The high resolution spectroscopy of X-Ray Imaging and Spectroscopy Mission \citep[XRISM;][]{Tashiro_2020} with its microcalorimeter Resolve \citep{Ishisaki_2018} will enable us to observe the line profile of iron emission lines and distinguish them.


\begin{acknowledgments}
\textit{NICER} analysis software and data calibration were provided by the NASA \textit{NICER} mission and the Astrophysics Explorers Program.
We thank K. Shibata (Doshisha University), H. Maehara (NAOJ) and K. Namekata (Kyoto University) for their useful comments and discussions.
We thank C. Markward (NASA/GSFC) for the calibration comments.
This research is supported by JSPS KAKENHI grant No. 24KJ1483 (S.I.).
T.E. was supported by the RIKEN Hakubi project.
Y.N. acknowledge support from NASA ADAP award program No. 80NSSC21K0632.
\end{acknowledgments}

\vspace{5mm}
\facilities{\textit{NICER} \citep{2016SPIE.9905E..1HG}}


\software{\texttt{astropy} \citep{astropy:2013, astropy:2018}, \texttt{HEAsoft}, \texttt{Xspec} \citep{1996ASPC..101...17A}, \texttt{PyXspec} \citep{Gordon_2021}, \texttt{PyAtomDB} \citep{Foster_2020}
          }

\appendix
\restartappendixnumbering 
 \section{Fine structure of the Fe XXV He$\alpha$ and XXVI Ly$\alpha$ lines} \label{appendix: Energy_table}
  \begin{table*}[b]
 \caption{Line center energy of the Fe XXV He$\alpha$ and Ly$\alpha$ lines}
\begin{center}
\begin{tabular}{cccc}
\hline \hline 
Ion               & $E_{l}^{NICER}$                 & Line Id. & Line Energy \\
                  & (keV)             &          & (keV)       \\ \hline
\multirow{4}{*}{Fe XXV} & \multirow{4}{*}{$\sim 6.69$} & w        &    6.700         \\
                  &                   & x        &    6.682         \\
                  &                   & y        &    6.668        \\
                  &                   & z        &     6.637        \\ \hline
\multirow{2}{*}{Fe XXVI} & \multirow{2}{*}{$\sim 6.97$} & Ly$\alpha_{1}$      &     6.973        \\
                  &                   & Ly$\alpha_{2}$      &         6.952   \\ \hline \hline
\end{tabular}
\end{center}
\tablecomments{$E_{l}^{NICER}$ means the line center energy in the \textit{NICER} spectra after convolving the detector response. The values of line energy are taken from \citet{Bianchi_2005, Wargelin_2005}.}
\label{tab:Energy_table}
\end{table*}
 
We display the line center energy of all lines of the Fe XXV He$\alpha$ and Ly$\alpha$ lines in Table \ref{tab:Energy_table} \citep[c.f. Table 1 of][]{Bianchi_2005}. As mentioned in Section \ref{sec:analysis_results}, these fine structures are not resolved in the \textit{NICER} spectra
 
  \color{black}
\section{Wideband spectra} \label{appendix: wideband}
\epsscale{1.15}
\begin{figure*}[]
\begin{center}
  \plotone{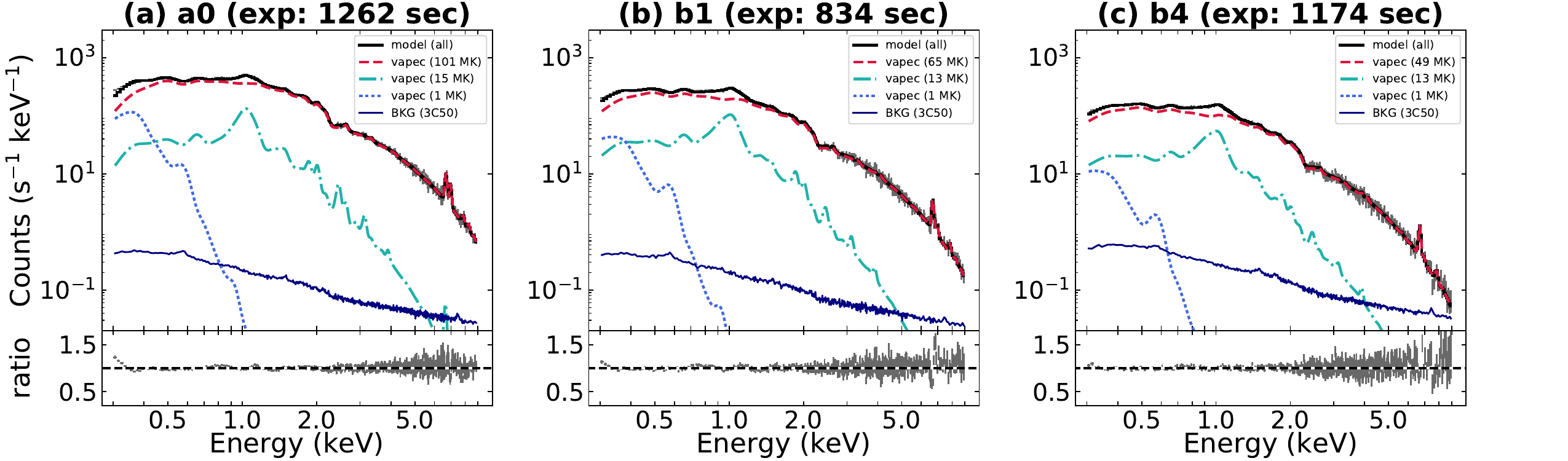}
\end{center}
  \caption{
  Same as Figure \ref{fig:wideband}, but during the interval a0, b1, and b4 for panels (a), (b), and (c), respectively.
  The temperatures of three components are 101 MK, 15MK, and 1 MK for panel (a), 65 MK, 13 MK, and 1 MK for panel (b), 49 MK, 13 MK, and 1 MK for panel (c).
  \label{fig:wideband_three}}
\end{figure*}

We provide wideband (0.3$-$9.0 keV) spectra of the interval a0, b1, and b4 fitted with the three temperature collisionally-ionized equilibrium (CIE) models (\texttt{vapec}) as examples in Figure \ref{fig:wideband_three}.
The interval a0 spectrum with the best photon statistics showed the slight systematic residuals around iron emission lines, which were not as much as the interval b1.
The interval b1 spectrum showed the most pronounced residuals around iron emission lines as discussed in Section \ref{sec:analysis_results}.
The interval b4 spectrum with the relatively poor photon statistics showed no systematic residuals around iron emission lines.
This is consistent with the fact that the Doppler velocity of the interval b4 spectrum obtained by our narrow band spectral analysis (Figure \ref{fig:parameter}e) was not blue-shifted.

\section{Fitting the interval b1 spectrum with Plasma models}\label{appendix: plasma_models}
\epsscale{0.90}
\begin{figure*}[]
  \plotone{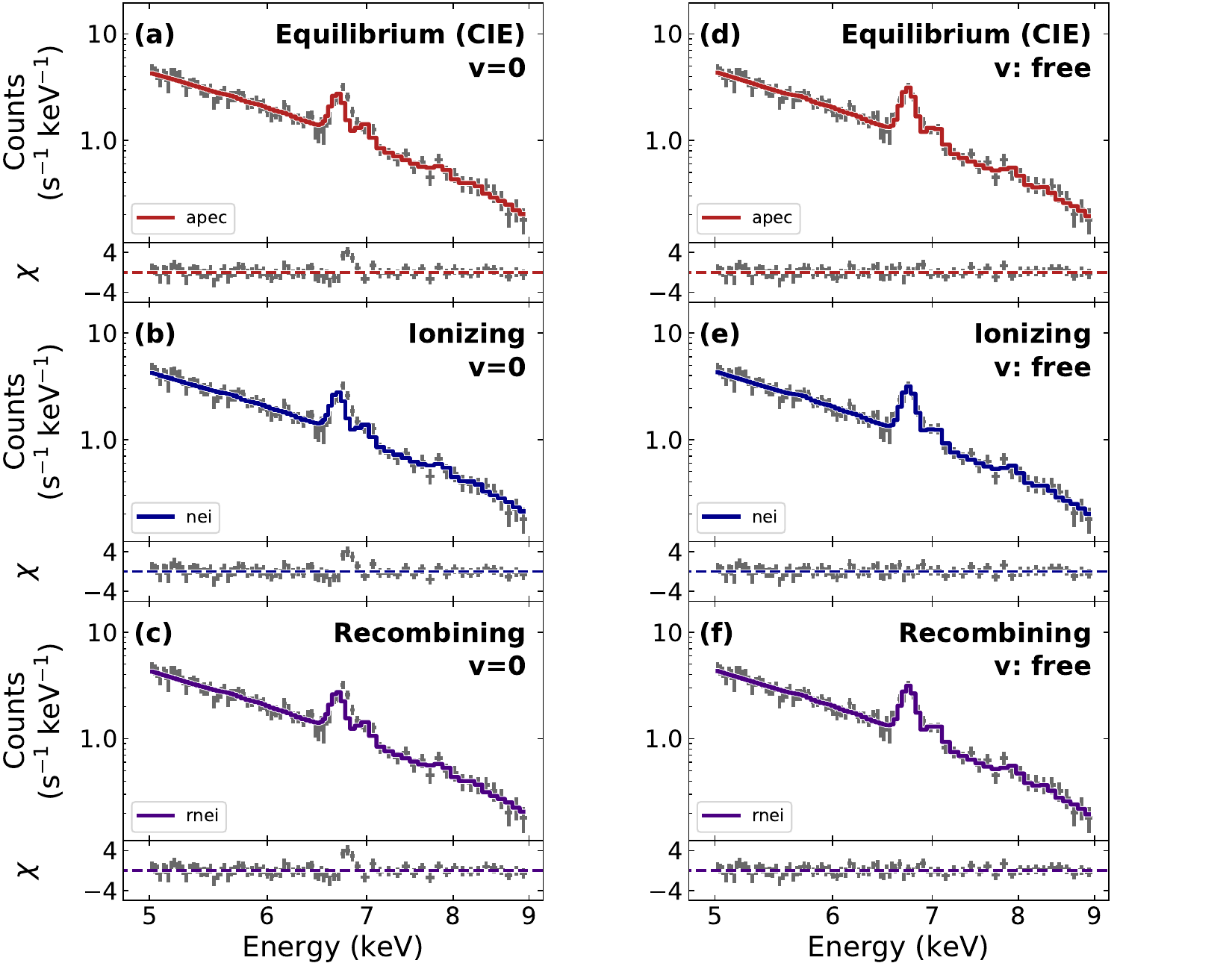}
  \caption{Comparison of spectral fittings by different plasma models around the iron emission lines during the interval b1: the CIE (\texttt{apec}, panel a), ionizing (\texttt{nei}, panel b), and recombining (\texttt{rnei}, panel c) plasma models with the velocity fixed at $v=0$. The lower panels show the fitting residuals.
  Panels d, e, and f are the same as panels a, b, and c, respectively, but letting the velocity of the plasma free.  
  \label{fig:spectra_appendix}}
\end{figure*}
\setcounter{table}{0}
\begin{table*}[t]
\caption{Best-fit spectral parameters of plasma models during the interval b1}
\begin{center}
\begin{tabular}{cccccc}
\hline \hline
\multicolumn{6}{c}{Collisional ionization equilibrium plasma model (\texttt{apec})}                                                                    \\ \hline \hline
\multicolumn{3}{c|}{Velocity: fixed at zero}  & \multicolumn{3}{c}{Velocity: free} \\ \hline 
\multirow{4}{*}{\texttt{apec}}                          & $kT$ (keV) / $T$ (MK)        & \multicolumn{1}{c|}{$6.90\pm0.69$ / $80.0\pm8.0$} & \multirow{4}{*}{\texttt{apec}}    & $kT$ (keV) / $T$ (MK)        & $6.16\pm0.65$ / $71.5\pm7.54$    \\
                                                        & $Z$ ($Z_{\odot}$) & \multicolumn{1}{c|}{$0.25\pm0.05$} &                                   & $Z$ ($Z_{\odot}$) & $0.28\pm0.05$ \\
                                                        & $v \: (10^{3} \: \mathrm{km \: s^{-1}})$               & \multicolumn{1}{c|}{$0.00$ (fix)} &                                   & $v \: (10^{3} \: \mathrm{km \: s^{-1}})$ & $-2.54\pm0.68$   \\
                                                        & $K^{\mathrm{apec}}$              & \multicolumn{1}{c|}{$0.49\pm0.04$} &                                   & $K^{\mathrm{apec}}$ & $0.51\pm0.05$    \\ \hline
\multicolumn{2}{c}{$\chi^{2}$ (d.o.f.)}                                     & \multicolumn{1}{c|}{86 (75)} & \multicolumn{2}{c}{$\chi^{2}$ (d.o.f.)} & 45 (74) \\
\multicolumn{2}{c}{Null hyp. prob.} & \multicolumn{1}{c|}{0.17} & \multicolumn{2}{c}{Null hyp. prob.} &   0.99  \\ \hline \hline
\multicolumn{6}{c}{Ionizing plasma model (\texttt{nei})}                                                                    \\ \hline \hline
\multicolumn{3}{c|}{Velocity: fixed at zero}  & \multicolumn{3}{c}{Velocity: free} \\ \hline 
\multirow{5}{*}{\texttt{nei}}                           & $kT$ (keV) / $T$ (MK) & \multicolumn{1}{c|}{$8.01\pm1.67$ / $92.9\pm19.4$} & \multirow{5}{*}{\texttt{nei}} & $kT$ (keV) / $T$ (MK) & $6.74\pm1.12$ / $78.2\pm13.0$\\
                                                        & $Z$ ($Z_{\odot}$)            & \multicolumn{1}{c|}{$0.24\pm0.06$} &                                   & $Z$ ($Z_{\odot}$) & $0.28\pm0.05$   \\
                                                        &$n_{e} t$ ($\mathrm{s\:cm^{-3}}$)& \multicolumn{1}{c|}{$\gtrsim 2 \times 10^{12}$} &                                   &$n_{e} t$ ($\mathrm{s\:cm^{-3}}$)& $\gtrsim 2 \times 10^{11}$    \\
                                                        & $v \: (10^{3} \: \mathrm{km \: s^{-1}})$               & \multicolumn{1}{c|}{$0.00$ (fix)} &                                   & $v \: (10^{3} \: \mathrm{km \: s^{-1}})$               & $-2.45\pm0.46$   \\
                                                        & $K^{\mathrm{nei}}$              & \multicolumn{1}{c|}{$0.46\pm0.06$} &                                   & $K^{\mathrm{nei}}$              & $0.49\pm0.06$   \\ \hline
\multicolumn{2}{c}{$\chi^{2}$ (d.o.f.)} & \multicolumn{1}{c|}{86 (74)} & \multicolumn{2}{c}{$\chi^{2}$ (d.o.f.)} &   44 (73) \\ 
\multicolumn{2}{c}{Null hyp. prob.} & \multicolumn{1}{c|}{0.16} & \multicolumn{2}{c}{Null hyp. prob.} &   0.99 \\ \hline \hline
\multicolumn{6}{c}{Recombining plasma model (\texttt{rnei})}                                                                    \\ \hline \hline
\multicolumn{3}{c|}{Velocity: fixed at zero}  & \multicolumn{3}{c}{Velocity: free} \\ \hline 
\multirow{6}{*}{\texttt{rnei}}                          & $kT$ (keV) / $T$ (MK)        & \multicolumn{1}{c|}{$7.00\pm0.76$ / $81.2\pm8.81$} & \multirow{6}{*}{\texttt{rnei}}    & $kT$ (keV) / $T$ (MK)        & $6.18\pm0.68$ / $71.7\pm7.89$   \\
                                                        & $kT_{\mathrm{init}}$ (keV) / $T_{\mathrm{init}}$ (MK)        & \multicolumn{1}{c|}{$9.06$ / $105$} & & $kT_{\mathrm{init}}$ (keV) / $T_{\mathrm{init}}$ (MK)       & 9.06 / 105    \\
                                                        & $Z$ ($Z_{\odot}$) & \multicolumn{1}{c|}{$0.25\pm0.05$} &                                   & $Z$ ($Z_{\odot}$) & $0.29\pm0.05$    \\
                                                        &$n_{e} t$ ($\mathrm{s\:cm^{-3}}$)& \multicolumn{1}{c|}{$\gtrsim 5 \times 10^{11}$} &  &$n_{e} t$ ($\mathrm{s\:cm^{-3}}$)& $\gtrsim 6 \times 10^{12}$   \\
                                                        & $v \: (10^{3} \: \mathrm{km \: s^{-1}})$               & \multicolumn{1}{c|}{0.00 (fix)} &                                   &    $v \: (10^{3} \: \mathrm{km \: s^{-1}})$               & $-2.63\pm0.53$    \\
                                                        & $K^{\mathrm{rnei}}$              & \multicolumn{1}{c|}{$0.49\pm0.04$} &                                   & $K^{\mathrm{rnei}}$              & $0.51\pm0.03$    \\ \hline
\multicolumn{2}{c}{$\chi^{2}$ (d.o.f.)} & \multicolumn{1}{c|}{87 (74)} & \multicolumn{2}{c}{$\chi^{2}$ (d.o.f.)} &  44 (73)  \\ 
\multicolumn{2}{c}{Null hyp. prob.} & \multicolumn{1}{c|}{0.15} & \multicolumn{2}{c}{Null hyp. prob.} &   0.99 \\ 
\hline \hline
\end{tabular}
\end{center}
\tablecomments{The lower limits of $n_{e} t$ in \texttt{nei} and \texttt{rnei} models are evaluated from the two-dimensional contour plots of ($kT$, $n_{e}t$) as the 90 \% confidence interval.
$K^{\mathrm{apec}}$, $K^{\mathrm{nei}}$, and $K^{\mathrm{rnei}}$ mean $10^{-14} (4 \pi)^{-1} [D_{\mathrm{A}} (1+z)]^{-2}] \int n_{\mathrm{e}} n_{\mathrm{H}} dV$, where $D_{\mathrm{A}}$ is the angular diameter distance to the source, $z$ is the redshift, $n_{\mathrm{e}}$ and $n_{\mathrm{H}}$ are the electron and hydrogen densities, and $dV$ is the volume element.}
\label{tab:fitted_parameter_appendix}
\end{table*}

We also tried CIE (\texttt{apec}) and off-CIE plasma models, the ionizing (\texttt{nei}) and recombining (\texttt{rnei}) models, for the interval b1 spectrum.
When we used the recombinning model, we fixed the initial temperature to the electron temperature of the interval a0 of $\sim 105$ MK.
All models can not eliminate the residuals at $v=0$ (Figure \ref{fig:spectra_appendix}a$-$c).
On the other hand, when we leave the velocity $v$ free and again fit the spectrum with \texttt{apec}, \texttt{nei}, and \texttt{rnei} (Figure \ref{fig:spectra_appendix}d$-$f), bulk motion, and thus the blue-shifted line center corresponding with the negative sign of the velocity, improved the fittings with better $p$-values, eliminating the residuals (Table \ref{tab:fitted_parameter_appendix}).
The best-fit parameters of the velocity were $-2540\pm680 \: \mathrm{km \: s^{-1}}$, $-2450\pm460 \: \mathrm{km \: s^{-1}}$, and $-2630\pm530 \: \mathrm{km \: s^{-1}}$ for \texttt{apec},  \texttt{nei}, and \texttt{rnei}, respectively.
These values are consistent with the velocity of $-2200 \pm 600 \: \mathrm{km \: s^{-1}}$ obtained with the phenomenological fitting (Section \ref{sec:analysis_results}) within error bars.
Furthermore, the best-fit parameters of temperature, abundance, velocity, and normalization became consistent within errorbars among the three plasma models.
This suggests that the observed 5$-$9 keV spectra can be explained either by CIE or off-CIE plasma models, assuming the plasma moving at a velocity of about $-2500$ $\mathrm{km \: s^{-1}}$. In other words, no model can reproduce the observed iron emission lines without the blueshifts.

\section{Energy determination accuracy of \textit{NICER} around iron emission lines} \label{appendix: nicer_accuracy}
\epsscale{0.99}
\setcounter{figure}{0}
\begin{figure*}[]
\begin{center}
  \plotone{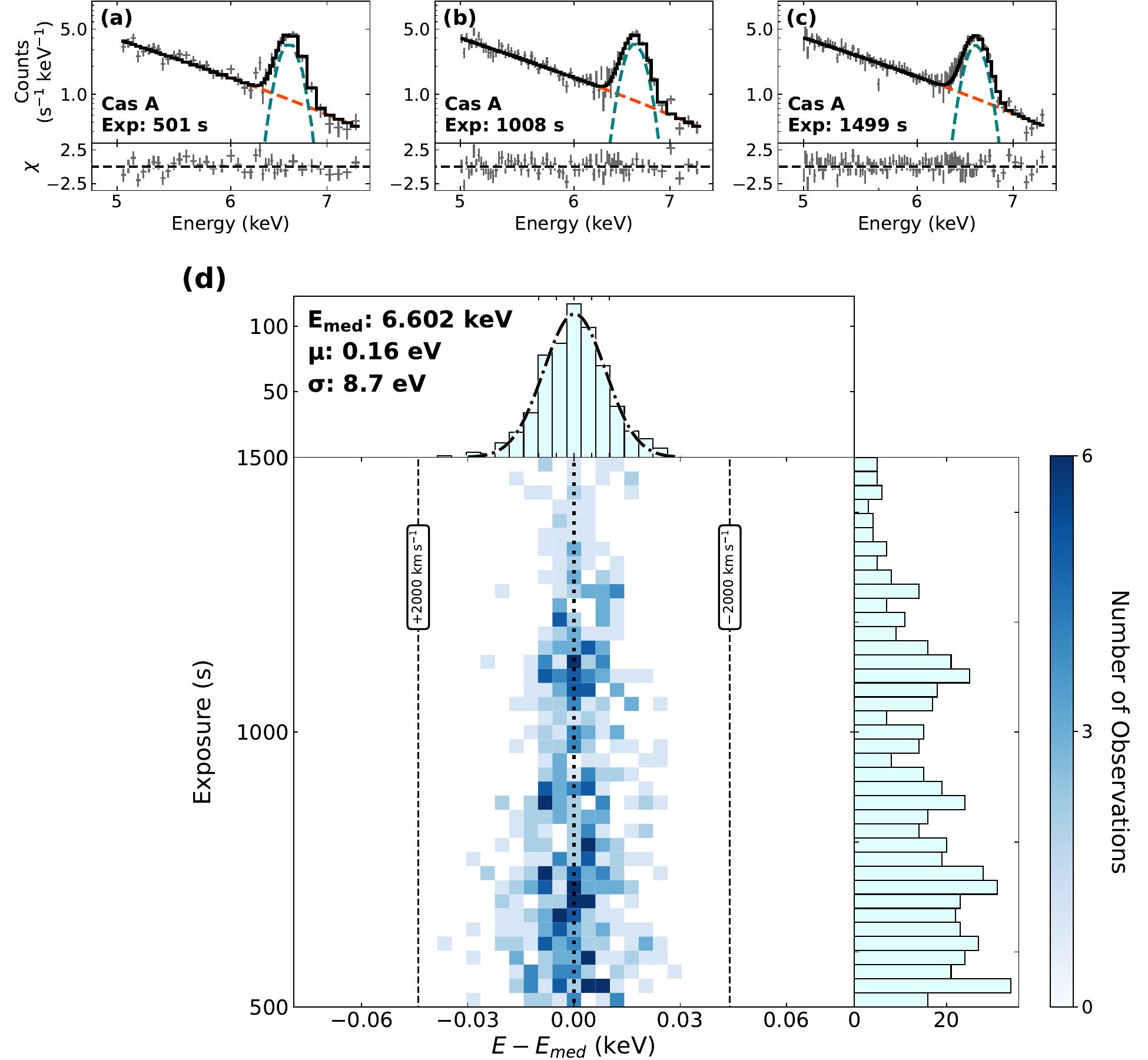}
\end{center}
  \caption{(a)$-$(c) Examples of \textit{NICER} Cas A spectra fitted with the phenomenological model of \texttt{powerlaw+gauss}, whose exposure times are (a) $\sim 500$ sec, (b) $\sim 1000$ sec, and (c) $\sim 1500$ sec, respectively. Orange and green dashed lines indicate \texttt{powerlaw} and \texttt{gauss} components, respectively.
  (d) The distribution of the line center of the gaussian function ($\Delta E = E - E_{\mathrm{med}}$). The upper panel shows the histogram of the distribution for all data ($N = 605$), and the Gaussian fitting (black dashdot line). The color map shows the distribution at each range of the exposure time. The right panel shows the histogram of the distribution at each exposure time.
  \label{fig:histgram}}
\end{figure*}

As discussed in Section \ref{sec:analysis_results}, the  Fe XXV He$\mathrm{\alpha}$ line was seem to be blueshifted during the flare.
However, the accuracy of the \textit{NICER} response function in determining the energy of photons should be checked.
Therefore, we need to investigate how uncertainty the \textit{NICER} response function has and whether the observed blueshifts are attributed to the astronomical phenomenon.

We analyzed a large number of \textit{NICER} data (192 ObsIDs) of the supernova remnant Cas A \citep[e.g.,][]{Holt_1994, Maeda_2009}, which has a count rate comparable to that of IM Peg during the flare, the strong iron emission line, and no time variability.
We downloaded all Cas A data observed by \textit{NICER} from HEASARC archive and processed them in the same manner as described in Section \ref{sec:observation_reduction}.
Then, we extracted GTI-divided and all-GTI spectra from all Obs-ID data.
As shown in Figure \ref{fig:histgram}a$-$c, we fitted the spectra, whose exposure times are $500 < t_{\mathrm{exp}} < 1500$ sec, in 5$-$7.4 keV with the phenomenological model (\texttt{powerlaw+gauss}).
The exposure time of most intervals during the flare on IM Peg is in this range.
The total number of Cas A spectra we fitted was 605.

Figure \ref{fig:histgram}d shows the distribution of the line center of the Gaussian.
The median value of the line center was $E_{\mathrm{med}} = 6.602$ keV.
The central value and standard deviation of $\Delta E$ $(= E - E_{\mathrm{med}})$ were $\mu = (1.6 \pm 3.1 )\times10^{-4}$ keV and $\sigma = (8.7\pm0.3) \times 10^{-3}$ keV, respectively.
The error ranges of $\mu$ and $\sigma$ are 90 \% confidence level.
\citet{Holt_1994} reported the difference in the Doppler shift between the locations in Cas A.
According to their observations with ASCA Observatory, the range of the line center energy of the iron emission line is $6.5568-6.6080$ keV \citep[see Table 1 in][]{Holt_1994}.
On the other hand, \citet{Maeda_2009} showed the line center energy is $6.621$ keV from the spectrum extracted from the entire Cas A observed with Suzaku.
\textit{NICER} 90 \% confidence level of the line center in our analysis ($6.60185 \: \mathrm{keV} < E_{\mathrm{med}} + \mu < 6.60247 \: \mathrm{keV}$) is in the range of these previous works.
Furthermore, our results indicates that the upper limit of the uncertainty of the \textit{NICER} response function is limited to $|\Delta E| \lesssim 3 \sigma = 2.6 \times 10^{-2}$ keV, which corresponds to the Doppler velocity $v \sim \pm 1200 \: \mathrm{km \: s^{-1}}$.
This means that the $-2200 \pm 600 \: \mathrm{km \: s^{-1}}$ ($(5 \pm 1) \times 10^{-2}$ keV) blueshift detected during the interval b1 ($t_{\mathrm{exp}} = 834$ sec) can not be explained by the uncertainty of the \textit{NICER} energy determination.

Furthermore, since we were multiplying both modeling and observed spectra by the response files at each time and taking the difference of the line center between them (Section \ref{sec:analysis_results}), the response-derived Doppler shift should cancel out to some extent.
Given these considerations, it is highly possible that the astronomical phenomenon made the observed blueshifts.

\section{Fitting the Fe XXV He$\alpha$ line with two gaussian components} \label{appendix: two_gaussian}
\setcounter{figure}{0}
\begin{figure*}[]
\begin{center}
  \plotone{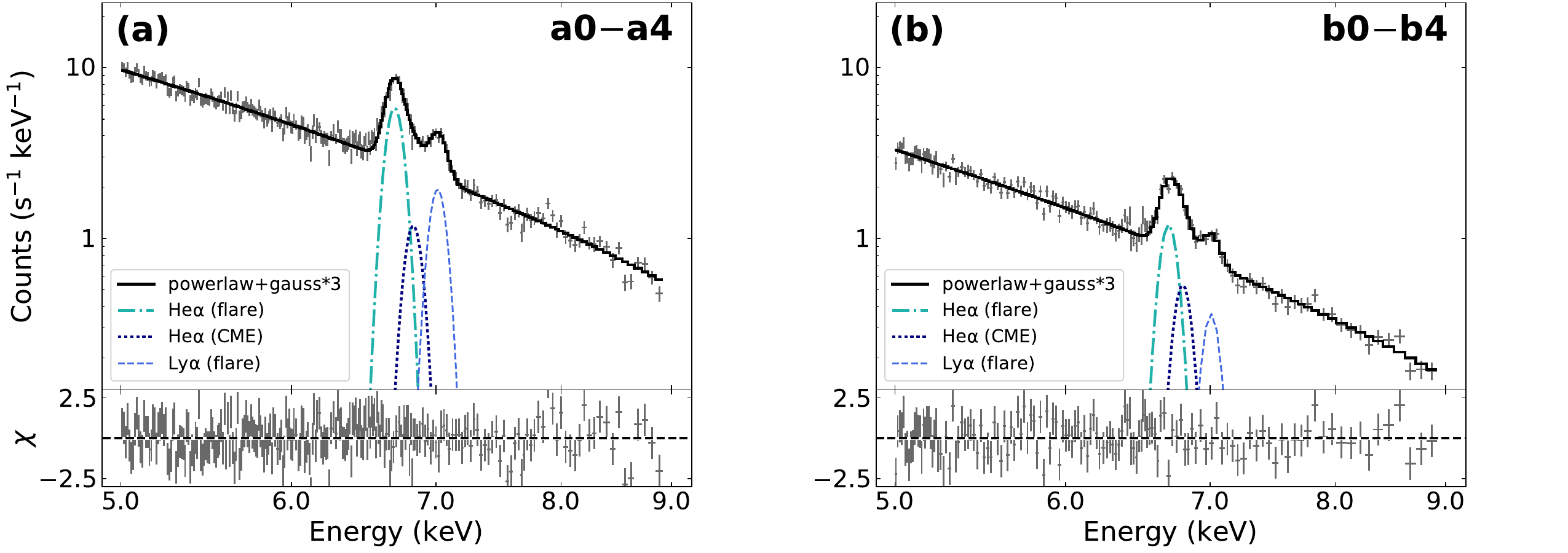}
\end{center}
  \caption{Background-subtracted 5$-$9 keV spectra of IM Peg fitted with a phenomenological model (\texttt{powerlaw+gauss*3}) during the interval a0$-$a4 and b0$-$b4 for panels (a) and (b), respectively.
  Green dashdot and blue dashed lines correspond to the Fe XXV He$\alpha$ and the Fe XXVI Ly$\alpha$ lines, respectively, whose line centers are fixed to the values in the laboratory frame.
  Navy dotted lines indicate the blue-shifted components of the Fe XXV He$\alpha$ line.
  \label{fig:two_gaussian}}
\end{figure*}

\epsscale{0.79}
\begin{figure*}[]
\begin{center}
  \plotone{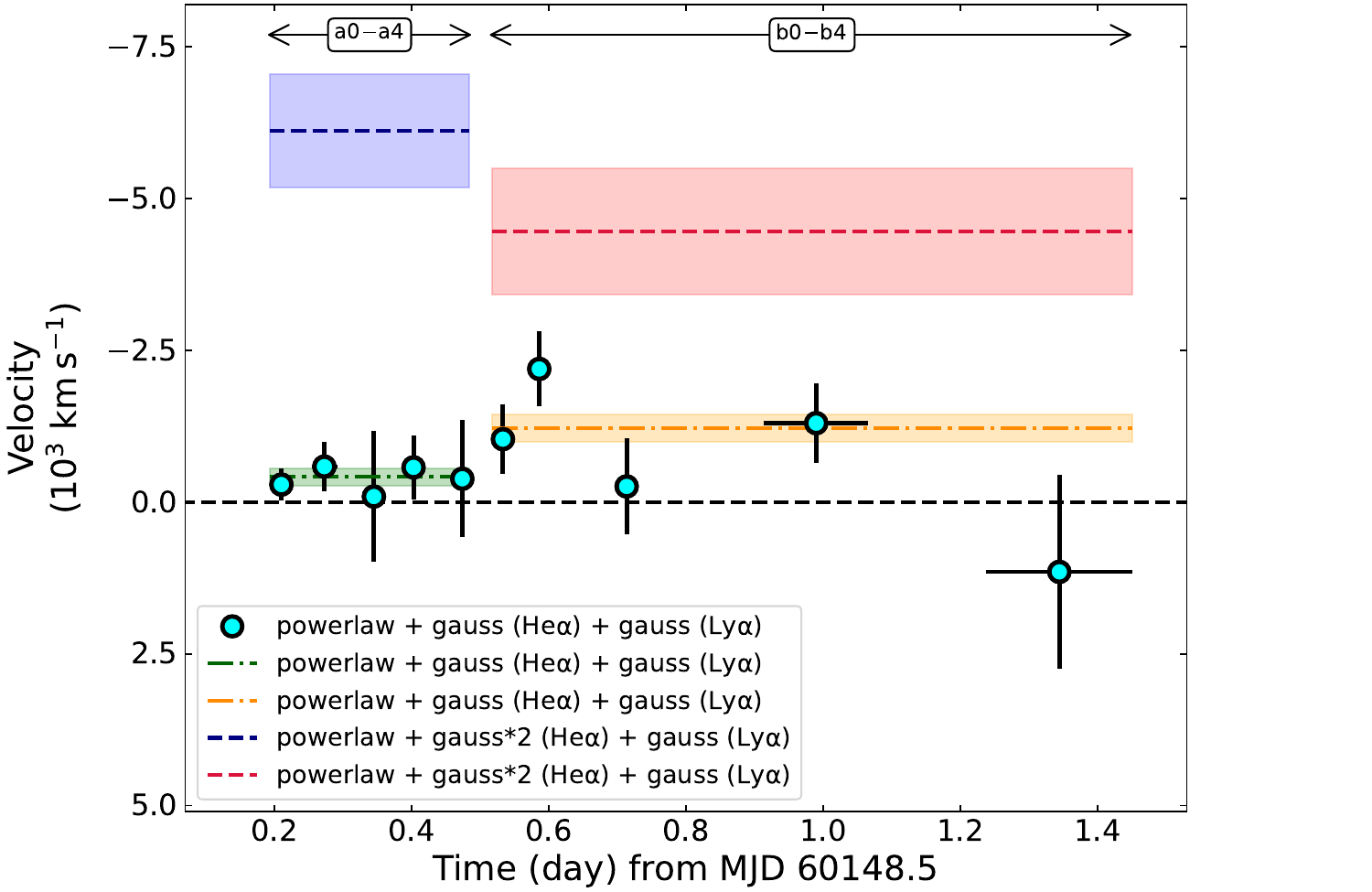}
\end{center}
  \caption{Doppler velocity of the blue-shifted components of the Fe XXV He$\alpha$ line (see e.g., Figure \ref{fig:two_gaussian}) overlaid on that of the line center of the Fe XXV He$\alpha$ line shown in Figure \ref{fig:parameter}g.
  Navy and red dashed lines indicate the Doppler velocity of the blue-shifted components of the Fe XXV He$\alpha$ line with the $90\%$ error range during the combined interval a0$-$a4 and b0$-$b4, respectively, when we fit the spectra with \texttt{powerlaw+gauss*3} model.
    Green and orange dashdot lines indicate the Doppler velocity of the line center of the Fe XXV He$\alpha$ line with the $90\%$ error range during the combined interval a0$-$a4 and b0$-$b4, respectively, when we fit the spectra with \texttt{powerlaw+gauss*2} model.
  \label{fig:velocity_for_all_phase}}
\end{figure*}

When we consider the blueshifts of the Fe XXV He$\alpha$ line being attributed to a CME, it is physically reasonable that there are both non-shifted (flare) and blue-shifted (CME) components of the Fe XXV He$\alpha$ line as discussed in the case of the blue-shifted H$\alpha$ line \citep[e.g.,][]{Inoue_2023}.
Then, we also fitted the interval a0$-$a4 and b0$-$b4 spectra with the phenomenological model (\texttt{powerlaw+gauss*3}) considering the CME components (Figure \ref{fig:two_gaussian}).
We did not fit the individual interval spectra with this model because photon statistics are insufficient to determine the CME components.
We fixed the line centers of the flare components of the Fe XXV He$\alpha$ and Fe XXVI Ly$\alpha$ lines to the lab-frame ones. 
As a result, the best-fit values of the line centers of the CME components of the Fe XXV He$\alpha$ line were $6.83 \pm 0.02$ keV and $6.79 \pm 0.02$ keV during interval a0$-$a4 and b0$-$b4, respectively.
These values correspond to the Doppler velocities of $-6100 \pm 900$ $\mathrm{km \: s^{-1}}$ and $-4400 \pm 1000$ $\mathrm{km \: s^{-1}}$, respectively (Figure \ref{fig:velocity_for_all_phase}).

\color{black}

\bibliography{sample631}{}
\bibliographystyle{aasjournal}


\end{document}